\documentclass[aps,prl,reprint,amsmath,amssymb,nofootinbib,superscriptaddress]{revtex4-1}
\usepackage{custom}
\usepackage{bm}

\begin{document}

\title{Maximum entropy freezeout of hydrodynamic fluctuations}

\author{Maneesha Sushama Pradeep}
\author{Mikhail Stephanov}

\affiliation{Department of Physics, University of Illinois, Chicago,
  IL 60607, USA}

\date{\today}

\begin{abstract}
  We propose a general approach to freezing out {\em fluctuations} in
  heavy-ion collisions using the
  principle of maximum entropy. We find the results naturally
  expressed as a direct relationship between the 
  {\em irreducible
    relative correlators} quantifying the deviations of hydrodynamic
   as well as hadron gas fluctuations from the ideal hadron gas
   baseline. The method also allows us to determine
   heretofore unknown parameters crucial for the freezeout of
   fluctuations near the QCD critical point in terms of the QCD
   equation of state.
\end{abstract}

\maketitle

\section{Introduction}
\label{sec:intro}
Mapping the phase diagram of QCD is the primary goal of heavy-ion
collision experiments~\cite{STAR:2010vob,Bzdak:2019pkr}.  Fluctuations
carry important information about the phase diagram. In particular,
non-monotonic behavior of fluctuation measures is a signature of the
QCD critical
point~\cite{Stephanov:1998dy,Stephanov:1999zu,Stephanov:2008qz,Athanasiou:2010kw,Stephanov:2011pb}.
Fluctuations are also important for the study of hydrodynamics in
small systems as well as understanding of initial conditions
\cite{Nagle:2018nvi}. Non-equilibrium dynamics of fluctuations is
important in general, and of particular importance is the dynamics of
the {\em non-Gaussian} fluctuations near the critical point. For many
of these reasons evolution of fluctuations in hydrodynamics has been
intensively studied
recently~\cite{Kapusta:2011gt,Ling:2013ksb,Stephanov:2017ghc,Nahrgang:2018afz,Nahrgang:2018afz,Akamatsu:2017,Akamatsu:2018,Akamatsu:2018tcp,Martinez:2017fbb,Martinez:2018,An:2019csj,An:2019osr,Rajagopal:2019xwg,Du:2020bxp,An:2020vri,Sogabe:2021svv,Pradeep:2022mkf}.

Freezeout (particlization) is the crucial step in translating
hydrodynamic fluctuations into fluctuations and correlations of
particles observed in experiment. There have been attempts to develop
appropriate procedure for fluctuation freezeout, starting from
Ref.\cite{Kapusta:2011gt}. One of the known problems is the proper
separation of the trivial (single-particle) contribution from the
multi-particle fluctuations induced by hydrodynamic correlations. This
problem was addressed in \cite{Ling:2013ksb,De:2020yyx} for one
particular case of charge diffusion, but a general solution is still
lacking, especially for non-Gaussian fluctuations. Indeed, freezeout
of non-Gaussian {\em out-of-equilibrium} fluctuations have not yet been
adequately addressed, except for a proposal in
Ref.~\cite{Pradeep:2022mkf}, where the procedure
introduced earlier for equilibrium fluctuations was straightforwardly
generalized in a somewhat ad hoc manner. Even in this case, the
proposal is limited to leading (most singular) critical contribution
to fluctuations. While fluctuating hydrodynamics itself obeys conservation laws, implementing these laws at freezeout is a nontrivial issue~\cite{Oliinychenko:2019zfk}.

In this Letter we introduce a very general method of freezing out
fluctuations based on the principle of maximum entropy. In this
approach the fluctuations of conserved quantities are matched exactly
(i.e., not only the leading critical contribution) between the
hydrodynamic and the kinetic (particlized) side of the freezeout
transition as dictated by conservation laws. The distribution of
fluctuations across particle momenta is then determined by maximizing
the entropy of fluctuations -- an object we introduce in this paper,
which is mathematically similar to the $n$-PI action of the quantum
field theory. The entropy of fluctuations has also a lot in common
with the concept of relative entropy, which, in this case, measures
the amount of entropy deficit of a system with known correlations
(relative to the most agnostic state corresponding to complete thermal
equilibrium).

Of course, freezeout procedure, being essential for translating
hydrodynamics into particle observables, has been known for a long
time in the form introduced by Cooper and
Frye~\cite{Cooper:1974mv}. This procedure, however, could only deal
with {\em mean} hydrodynamic quantities, i.e., single-particle
observables. It does not address the question of how to freeze
out the {\em fluctuating} hydrodynamics.

In this Letter we apply the general principle of maximum entropy to
the long-standing problem of the fluctuation freezeout and obtain
several novel results, which non-trivially match some of the existing
approaches to freezeout in the literature, while augmenting or
correcting others. Most importantly, the new approach allows us to
tackle the problem of the freezeout of non-Gaussian fluctuations, in
and out of equilibrium.  Furthermore we are now able to determine some
of the thus far unknown parameters crucial for the freezeout of the
fluctuations near the critical point.

\section{Setup and main result}
\label{sec:setup}
The principle of maximum entropy has been applied recently to
implement freezeout of mean hydrodynamic variables into
single-particle observables in Ref.~\cite{Everett:2021ulz}.  The
procedure reproduces the Cooper-Frye procedure when 
non-equilibrium effects are ignored, but allows systematic
incorporation of non-equilibrium, dissipative effects. Let us,
therefore, begin by briefly reviewing the maximum entropy freezeout in the
simplest case of mean quantities in equilibrium.

The aim is to find
the phase space distribution function $f_A\equiv f_{\tilde A}(x_A)$,
where by $A$ we denote a composite index describing discrete quantum
numbers of particles such as spin or baryon number ($q_A$) as well
their momenta $p_A$ (collectively $\tilde A$) and the coordinate
$x_A$. In other words, $A$ labels a ``cell'' in the phase space. In
position space this cell matches a hydrodynamic cell, which has a
small but macroscopic size. The function $f_A$ is the mean occupation
number of the available single-particle states in the cell~$A$.  It must be
such that the conserved hydrodynamic quantities, such as
energy-momentum and conserved charge (such as baryon number) densities
in the rest frame of the fluid, $T^{\mu\nu}u_\nu = \epsilon u^\mu$ and
$J\cdot u = n$, are matched locally at each point $x$ (i.e., in each
cell) on the freezeout hypersurface by the gas of non-interacting
hadrons (resonance gas), i.e.,
\begin{equation}\label{eq:en-matching}
    \epsilon u^{\mu}(x)=\int_{\tilde A} \,
      p_A^{\mu}f_{\tilde A}(x)\,,\quad
    n(x)=\int_{\tilde A}\, q_A f_{\tilde A}(x)\,,
\end{equation}
where $\int_{\tilde A}$ denotes the summation over particle species $A$ as
well as the integration over their momenta (with Lorentz invariant measure).
Obviously, there are infinitely many solutions to the constraints
(\ref{eq:en-matching}) on $f_A$. We expect that the most likely
solution to describe the distribution of particles after freezeout
maximizes the entropy of the ideal hadron (resonance) gas given by
\begin{equation}
  \label{eq:S1-sum}
  S[f]=\int_x \int_{\tilde A} \,S_A\,,
\end{equation}
with
\begin{equation}
  \label{eq:S_A}
  S_A= \frac{1+\theta_Af_A}{\theta_A}\log (1+\theta_Af_A)-f_A \log f_A\,,
\end{equation}
where $\theta_A$ is $\pm1$ for Bose/Fermi particles (or 0 for
classical particles) and $\int_{x}$ is the 3-volume integration over
the freezeout hypersurface.  Introducing Lagrange multipliers
$\beta_\mu$ and $-\alpha$ for quantities in
Eqs.~(\ref{eq:en-matching}) and solving the variational problem we
find $f_A=(e^{\beta\cdot p_A - \alpha q_A}-\theta_A)^{-1}$, where $\beta_\mu$
and $-\alpha$ are determined by solving Eqs.~(\ref{eq:en-matching})
and correspond to the temperature $T=(-\beta\cdot u)^{-1}$ and
chemical potential $\mu=T\alpha$ of the hadron gas for given
values of energy-momentum and baryon densities in
Eq.~(\ref{eq:en-matching}). Thus, the maximum entropy approach
reproduces the standard Cooper-Frye prescription. As shown in
Ref.~\cite{Everett:2021ulz} this approach can be
naturally extended
to incorporate matching to viscous stress and diffusive current in
hydrodynamics against the non-ideal corrections to particle distribution
function $f_A$, i.e., imposing additional constraints, but using the
same entropy $S[f]$.

In this Letter we show that a more general application of the
principle of maximum entropy can also provide a natural solution to
the problem of freezing out the fluctuating hydrodynamics. In this
case hydrodynamic fluctuations will be matched by fluctuations in the
hadron  gas, and the entropy must be a functional of the
measures of fluctuations, i.e., of the {\em correlators} of
particle distributions $f_A$.

To simplify
notations and to make them more general, we shall organize
hydrodynamic variables in a vector $\Psi^a$, where, e.g.,
$\Psi^\mu=\epsilon u^\mu$ and $\Psi^5=n$. We also denote by $P^a_A$
the contribution of a single particle in the phase space cell $A$ to the hydrodynamic variable
$\Psi^a$ in the hydrodynamic cell at point $x_a$, i.e., $P^\mu_A=p^\mu_A\delta^{3}(x_a-x_A)$ and $P^5_A=q_A\delta^{3}(x_a-x_A)$, so that
Eqs.~(\ref{eq:en-matching}) can be written as
$\Psi^a=\int_{A}P^a_A f_A$ and, correspondingly,
$\delta\Psi^a=\int_{A}P^a_A \delta f_A$, which translates
into a relationship between the (connected) correlators in
the hadron  gas given by
$G_{AB\ldots}=\langle\delta f_A\delta f_B\ldots\rangle_c$ and
(connected) correlators of hydrodynamic variables
$H^{ab\cdots}=\langle\delta\Psi^a\delta\Psi^b\ldots \rangle_c $:
\begin{equation}
  \label{eq:HG}
  H^{ab\ldots} = \int_{AB\ldots} G_{AB\ldots}P_A^aP_B^b\ldots\,,
\end{equation}
where $\int_A\equiv\int_{x_A}\int_{\tilde A}$. These $n$-point hydrodynamic
correlators ($H_n$ in shorthand) are related to $n$-point Wigner
functions $W_n$ 
by the generalized Wigner transform introduced in Ref.~\cite{An:2020vri}.

We treat Eqs.~(\ref{eq:HG}) as constraints to be obeyed by
$G_{AB\ldots}$ ($G_n$ in shorthand). The maximum entropy principle
will then determine $G_n$ by maximizing the entropy
$S[f,G,G_3,G_4\ldots]$ of the state of the hadron gas with given
fluctuations characterized by connected correlators~$G_n$. We
determine this entropy functional by a calculation similar to
Ref.~\cite{Stephanov:2017ghc} but for higher-order correlators.

Solving the variational problem for $G_n$ with constraints in
Eq.~(\ref{eq:HG}) we find relationships between the hydrodynamic,
$H_n$, and particle, $G_n$, correlators.  These relationships are
especially simple and intuitive to linear order in relative
correlators $\Delta G_n \equiv G_n-\bar G_n$ and
$\Delta H_n=H_n-\bar H_n$, expressing correlations relative to the
baseline $\bar G_n$ and $\bar H_n$ given by the ideal hadron gas in
equilibrium.  In order to express these relationships we define (see
Eq.~(\ref{eq:hatG})) a somewhat novel kind of correlation measures
$\widehat\Delta G_n$ (and similarly, $\widehat\Delta H_n$) which we
refer to as {\it irreducible relative correlators (IRC)}. These
measures, similarly to $\Delta G_n$, quantify correlations {\em
  relative} to the ideal resonance gas, but only correlations not
reducible to lower-order correlations.
For a classical gas ($\theta_A=0$)
$\widehat\Delta G_n$ are similar to irreducible correlators described in
Ref.~\cite{Ling:2015yau} or ``correlation functions'' in
Ref.~\cite{Bzdak:2017ltv}. In terms of the IRCs the relationships
translating hydrodynamic fluctuations into particle fluctuaions take
the following form:
\begin{equation}
  \label{eq:DhatGDhatH}
  \widehat\Delta G_{AB\dots}
  = \sum_{ab\dots}\int_{x_ax_b\dots} \!
 \widehat\Delta H_{ab\dots}
  (\bar H^{-1}P\bar G)^a_A (\bar H^{-1}P\bar G)^b_B\ldots
\end{equation}
where
$(\bar H^{-1}P\bar G)^a_A\equiv\sum_{a'}\int_{x_a'}\int_{A'}\bar
H^{-1}_{aa'}P^{a'}_{A'}\bar G_{A'A}$.  In practice, the integrals over
$x_{a'}$ and $A'$ are trivial since $P^{a'}_{A'}$ and $\bar G_{A'A}$ are both
delta-functions of their spatial coordinates and $\bar G_{A'A}$ is
also a delta-function of the momenta (and other particle quantum
numbers), i.e.
$\bar G_{A'A}= f'_{A} \delta_{A'A}\sim
\delta^3(x_{A'}-x_A)\delta^3(p_{A'}-p_{A})$, where $f'_A=f_A(1+\theta_Af_A)$.

In what follows we do not write explicitly, but imply,
the summation/integration corresponding to repeated indices labeling either
hydrodynamic variables (and cells) or hadron gas variables (and
phase-space cells).

\section{Entropy of fluctuations}
\label{sec:entropy-fluctuations}
Recall that the exponential of the entropy $S$ in
Eq.~(\ref{eq:S1-sum}) is proportional to the number of microstates of
the system with given values of occupation numbers $f_A$ of the
hydrodynamic cells. In the thermodynamic (large volume) limit there is
a large number of single-particle quantum states in each hydrodynamic
cell and the number of possible ways to occupy these elementary
quantum states is exponentially large, of order $e^S$. A macroscopic
state is an ensemble of this exponentially large number of microscopic
states with occupation numbers close to mean $f_A$. The values
$f_A$ are not the same in all microscopic states, but fluctuate. The
magnitude of fluctuations is suppressed in the thermodynamic
limit. The probability distribution of these fluctuations is given by
the exponential of $S$ in Eq.~(\ref{eq:S1-sum}) with additional
constrains given by Eq.~(\ref{eq:en-matching}). As usual, in
thermodynamic limit, one can implement these constraints using
Lagrange multipliers, i.e., using the probability distribution
$\exp(S+J_a\Psi_a)$ and choosing $J_a$ to satisfy the constraints on
$\Psi_a$. Using this probability distribution one can calculate the
expectation values $\bar G_{AB\ldots}$ of the fluctuation correlators
in equilibrium which will depend on $\Psi^a$.

We can also consider states with (some of) the correlators
$G_{AB\dots}$ having specific values, not necessarily equal to
$\bar G_{AB\dots}$. These states must have lower entropy since more
information is available about these states.  To find their entropy we
can consider the probability distribution perturbed by additional
factor $\exp(K_{AB\dots}f_Af_B\dots)$, where $K_{AB\dots}$ play the
role similar to Lagrange multipliers. Integrating over fluctuations of
$f_A$ we can then obtain $G_{AB\dots}$, which will depend on
$K_{AB\dots}$. Solving for $K$ and substituting back into the
probability distribution we find the probability distribution for
$f_A$ with given correlators $G_{AB\dots}$. The Gibbs entropy of this
probability distribution is the key quantity, which we would then
maximize subject to constraints on $G_{AB\dots}$ from Eq.~(\ref{eq:HG}).

The calculation of the entropy of fluctuations along these lines for a
two-point correlator can be found in
Ref.~\cite{Stephanov:2017ghc}, where it is also pointed out
that the result mathematically resembles the 2-PI action in quantum field theory~\cite{Luttinger:1960ua,Baym:1962sx,Cornwall:1974vz,Norton:1974bm,Berges:2003pc}:
\begin{equation}
  \label{eq:S2}
    S_2 = S
  + \frac12{\rm Tr\,}\left[
\log(- CG) + CG + 1
  \right]\,,
\end{equation}
where $C_{AB}\equiv\delta^2 S/(\delta f_A\delta f_B)$ and  $G\equiv G_2$. The difference
$S_2-S$ vanishes when $G$ equals $-C^{-1}\equiv \bar G$ and can be viewed as
the additional (negative) entropy of the state with additional
constraints on correlators relative to the entropy of the state with
correlations given simply by $\bar G$.

We want now to determine the correlator $G_{AB}$ satisfying the
constraints in Eq.~(\ref{eq:HG}).
The most likely value of $G$ is given by the maximum
of the entropy $S_2$ subject to these constraints. Introducing Lagrange
multiplier matrix $\Lambda_{ab}$ we find, solving the constrained
variational problem, that
\begin{equation}
  \label{eq:G-Lambda2}
  G^{-1}_{AB} = \bar G^{-1}_{AB} +\Lambda_{ab}\, P^{a}_{A}\, P^{b}_{B}\,.
\end{equation}
We can then determine the Lagrange multipliers by substituting
(\ref{eq:G-Lambda2}) into (\ref{eq:HG}) and we find
\begin{equation}
  \label{eq:Lambda}
  \Lambda=H^{-1}-\bar H^{-1}\,.
\end{equation}
Substituting into Eq.~(\ref{eq:G-Lambda2}) we  obtain
\begin{equation}
  G^{-1}_{AB} = \bar G^{-1}_{AB} + (H^{-1}-\bar H^{-1})_{ab}
\, P^{a}_{A}\, P^{b}_{B}\,.
\label{eq:G2}
\end{equation}

\section{Non-Gaussian fluctuations}
\label{sec:non-gauss-fluct}
Extending this calculation to higher-order correlators, specifically
to $n=3,4$ relevant for non-Gaussian fluctuations in experiments~\cite{STAR:2010vob,Bzdak:2019pkr,STAR:2021iop,STAR:2021fge}, we obtain the
entropy $S_4[f,G,G_3,G_4]$ as
  \begin{multline}
    \label{eq:S32}
    S_3 = S_2 \,+ \,:\!\frac{1}{6}C_{ABC}G_{ABC}
    +\frac{1}{8}C_{ABCD}G_{AC}G_{BD}\\
    -\frac{1}{12}G_{AB}^{-1} G_{CD}^{-1} G_{EF}^{-1} G_{ACE}G_{BDF}\!:\,,
  \end{multline}
 \begin{widetext}
 \begin{multline}
    \label{eq:S43}
    S_4 = S_3 \,  + \,:\!\frac{1}{24}C_{ABCD}G_{ABCD}
    -\frac{1}{48}G_{AC}^{-1} G_{BD}^{-1} G_{EF}^{-1} G_{HI}^{-1}
    G_{ABEH}G_{CDFI}\\
    +\frac{1}{8}G_{ABC}G_{DEF}G_{HIJK}G_{AH}^{-1}G_{BI}^{-1}G_{CF}^{-1}G_{DK}^{-1}G_{EJ}^{-1}
    -\frac{1}{16}G_{AB}^{-1}G^{-1}_{CD}G^{-1}_{EF}G^{-1}_{HI}G^{-1}_{JK}G^{-1}_{LM}G_{ACE}G_{FHJ}G_{IKL}G_{BDM}\\
    -\frac{1}{12}G_{AEH}G_{BCJ}G_{FDL}G_{IKM}G_{AB}^{-1}G^{-1}_{CD}G^{-1}_{EF}G^{-1}_{HI}G^{-1}_{JK}G^{-1}_{LM}\!:\,,
  \end{multline}
\end{widetext}
where $:\!X\!:\,\equiv X-\bar X$.  The result is mathematically similar to the
 $n$-PI action in QFT,\cite{Berges:2004pu} as was the case for $n=2$, with
$S_n$ corresponding to truncation at $n-1$ loops. (As discussed in
Ref.~\cite{An:2020vri} the loop expansion corresponds to expansion in the
magnitude of fluctuations.) We can now maximize the entropy
$S_4$ with respect to $G_3$ and $G_4$ subject to constraints from
hydrodynamic correlators $H_3$ and $H_4$ in Eq.~(\ref{eq:HG}) and find
\begin{multline}
  G_{ABC}=
  \Big[C_{QRS}+\Big\{\tp^{abc}
  - (PG)^a_T(PG)^b_U(PG)^c_V C_{TUV}\Big\}\\
  \times\(H^{-1}P\)^a_Q\(H^{-1}P\)^b_R\(H^{-1}P\)^c_S
\Big]
  G_{QA}G_{RB}G_{SC}\,;
  \label{eq:Gset13cm} 
\end{multline}
\begin{multline}\label{eq:G4}
  G_{ABCD}=
\Bigg(C_{QRST}
  +\Big[3G_{YX}C_{YQR} C_{XST}\Big]_{\overline{QRST}}\\
  +\Big\{\tp^{abcd}
  -    P^a_{I}P^b_{J}P^c_{K}P^d_{L}
  \Big(G_{IM}G_{JN}G_{KO}G_{LP}C_{MNOP}\\
    +3G^{-1}_{XY}\Big[G_{YIJ} G_{XKL}\Big]_{\overline{IJKL}}\Big)\Big\}\\
\times(H^{-1}P)^a_Q(H^{-1}P)^b_R(H^{-1}P)^c_S(H^{-1}P)^d_T
\Bigg)\\
 \times G_{QA}G_{RB}G_{SC}G_{TD}\,,
\end{multline}
where 
$C_{AB\dots}\equiv\delta^n S/(\delta f_A\delta f_B\dots)$ and we used the
notation $[\dots]_{\overline{ABC\dots}}$ for average over the
permutations of indices.

Equations (\ref{eq:G2}),~(\ref{eq:Gset13cm}) and~(\ref{eq:G4}) can be solved for correlators
$G_n$ iteratively. However, the structure of these equations is somewhat easier
to appreciate in the linearized limit, applicable  when the
correlations relative to hadron gas, i.e., $G_n-\bar G_n\equiv
\,:\!G_n\!:\,\equiv \Delta G_n$, are sufficiently small (a reasonable
approximation for heavy-ion collisions). In this limit the solution
can be expressed compactly by Eq.~(\ref{eq:DhatGDhatH}), or,
with summation/integration implied, as
\begin{equation}
  \label{eq:DhatGDhatH2}
  \widehat\Delta G_{AB\dots}
  =
  \widehat\Delta H_{ab\dots}
  (\bar H^{-1}P\bar G)^a_A (\bar H^{-1}P\bar G)^b_B\ldots,
\end{equation}
in terms of the
correlators $\widehat\Delta G_n$ and $\widehat\Delta H_n$,
which could be  termed {\em irreducible relative
(connected)  correlators (IRC)}. These correlators quantify ``genuine''
(i.e., not reducible to lower-order correlations) 
$n$-point correlations in $G_n$ {\em relative} to the ideal hadron gas $\bar
G_n$. This is achieved by
recursively subtracting these lower-order correlations:
\begin{multline}\label{eq:hatG}
   \widehat\Delta G_{AB} \equiv  \Delta G_{AB}\,;\\
    \widehat\Delta G_{ABC}\equiv \Big[\Delta G_{ABC}
    - 3\widehat\Delta G_{AD} (\bar G^{-1}\bar G_3)_{DBC}\Big]_{\overline{ABC}}\\ 
   \widehat\Delta G_{ABCD}  \equiv\Big[\Delta G_{ABCD}
    - 6 \widehat\Delta G_{ABF} (\bar G^{-1}\bar G_3)_{FCD}\,;
   \\
    -  4\widehat\Delta G_{AF}(\bar G^{-1}\bar G_4)_{FBCD}\\
    - 3\widehat\Delta G_{EF}(\bar G^{-1}\bar G_3)_{EAB}
    (\bar G^{-1}\bar G_3)_{FCD} \Big]_{\overline{ABCD}}\,.
\end{multline}
Similar relations define IRCs $\widehat\Delta H_n$  of
hydrodynamic variables, with $H$ instead of $G$ and indices $ab\dots$
instead of $AB\dots$.

Note that the factors
$(\bar G^{-1}\bar G_n)_{ABC\dots}\equiv \bar G^{-1}_{AX}\bar
G_{XBC\dots}$  in Eq.~(\ref{eq:hatG}), in the case of negligible
quantum statistics effects
(or $\theta_A=0$ in Eq.~(\ref{eq:S_A})), are equal to $\delta_{ABC\dots}$.
Thus, in this case, the IRCs
$\widehat\Delta G_n$ coincide with
correlators $C_{ab\dots}$ described in Ref.\cite{Ling:2015yau},
whose phase space integrals give factorial cumulants. Such correlators
and factorial cumulants play important role in the acceptance
dependence of the fluctuation
measures~\cite{Ling:2015yau,Bzdak:2017ltv}.

\section{Comparison with existing methods}
We can now compare the results of the maximum entropy approach with
other freezeout procedures used in the
literature to implement freezeout of fluctuations.

Ref.~\cite{Kapusta:2011gt} considered fluctuations of $f_A$ caused by
fluctuations of hydrodynamic parameters such as temperature and
chemical potential, $J_a$ in our notations, i.e.,
\begin{math}
\delta f_A = (\partial f_A/\partial J_a) \delta J_a=(P\bar G)^a_A\delta J_a\,,
\end{math}
where, as before, $\bar G_{AB}=f'_A\delta_{AB}$.
Using hydrodynamic correlators
$\langle\delta J_a\delta J_b\rangle= H^{-1}_{ab}  $
one then finds:
\begin{equation}
  \label{eq:ff}
  G_{AB} = H^{-1}_{ab} (P\bar G)^a_A(P\bar G)^b_B\,,
\end{equation}
as opposed to our Eq.~(\ref{eq:G2}).
We see that the problem with Eq.~(\ref{eq:ff}) is in the absence of the
separate contribution of the ideal gas fluctuations,
$\bar G_{AB}=f'_A\delta_{AB}$, which matches $\bar H$ in
hydrodynamics, but does not describe {\em correlations} between two
{\em different} particles \cite{Ling:2013ksb,De:2020yyx}. While the approach of
Ref.\cite{Kapusta:2011gt} could satisfy the constraints (\ref{eq:HG}),
it does so, in part, via spurious two-particle
correlations. This problem was addressed in
Ref.~\cite{Ling:2013ksb,De:2020yyx} for charge fluctuations,
where the ideal gas (Poisson) contribution to $H$ was subtracted
before applying ``freezeout (thermal) smearing'' to the remainder, $H-\bar H$ in our
notations. Thus, maximum entropy approach reproduces, in
Eq.~(\ref{eq:DhatGDhatH2}), the procedure in
Ref.~\cite{Ling:2013ksb,De:2020yyx} for two-point correlators. The
subtractions of lower order terms in Eqs.~(\ref{eq:hatG}) generalize
this procedure to higher-order correlators.

Fluctuations near the QCD critical point {\em in equilibrium} have
been described by considering a fluctuating critical mode $\sigma$
coupled to the observed particles via their $\sigma$-dependent
masses~Refs.~\cite{Stephanov:1998dy,Stephanov:1999zu,Stephanov:2008qz,Athanasiou:2010kw,Stephanov:2011pb,Bluhm:2016byc}.
This approach was further generalized in Ref.~\cite{Pradeep:2022mkf}
to {\em non-equilibrium } critical fluctuations by mapping the
correlators of $\sigma$ to correlators of the specific entropy
$m\equiv s/n$ -- the critical field in Hydro+
\cite{Stephanov:2017ghc}. We can now compare this approach to the
result of the maximum entropy method by considering only the matrix
element $H^{mm}$ of hydrodynamic correlator $H$ corresponding to the
fluctuations of the specific entropy $m$.

Furthermore, since this approach only
considers the leading (most singular) critical contribution, for our
comparison, we can
neglect lower-order correlations, which contribute subleading behavior
in terms of the dependence on the correlation length near the critical
point~\cite{Stephanov:2008qz}. In practice this means
$\widehat\Delta G_n=\Delta G_n$ up to subleading (less critical) terms.

Translating the freezeout prescription of Ref.~\cite{Pradeep:2022mkf} into
our notations we find:
\begin{equation}
  \label{eq:DG-h+fo}
   \Delta G_{AB}  =
   \frac{g_{A}g_{B}}{ZT^2}\frac{m_{A}}{E_{A}}\frac{m_{B}}{E_{B}}\Delta
   H^{\hs\hs} f'_Af'_B\,,  
 \end{equation}
where $g_{A,B}$ and $Z$ are
parameters describing the coupling of $\sigma$ to particles $A,B$
 (see Ref.~\cite{Pradeep:2022mkf}).
The maximum entropy freezeout gives
\begin{equation}
  \label{eq:DG-maxent2}
   \Delta G_{AB}  = \Delta H^{\hs\hs}
   (\bar H^{-1}P\bG)_{\hs A}(\bar H^{-1}P\bG)_{\hs B}\,.
\end{equation}
Comparing Eqs.~(\ref{eq:DG-h+fo}) and~(\ref{eq:DG-maxent2}) we find
that they could be reconciled if $g_A$ had particle energy
dependence given by the factor $P^m_A=\left(E_A-wq_A/n\right)/(nT)\delta^3(x_m-x_A)$
-- the contribution of
particle~$A$
to the fluctuation of $m=s/n$. The absence of the energy dependence of $g_A$ in Eq.~(\ref{eq:DG-h+fo}) is a
consequence of the simplifying assumption that the field $\sigma$
 couples to mass term.
Maximum entropy method allows us to relax
this assumption and determine the ``coupling'' $g_A$ together with its
energy dependence from the equation of state
(EOS) of QCD:
\begin{equation}\label{eq:ga2}
  g_{A} 
  =\sqrt{Z}\frac{E_A}{m_A}
  \(\bar H^{-1}\)_{\hs\hs}\frac{w_c}{n_c}\(\frac{E_{A}}{
      w_c}-\frac{q_A}{n_c}\)\,,
\end{equation}
where $(\bar H^{-1})_{mm}$ is the hadron gas contribution to the
fluctuations of specific entropy $m$, which can be also found from the
{\em non-singular} contribution to the EOS \cite{Parotto:2018pwx} as
$(\bar H^{-1})_{mm}=n^2/\bar c_p$.

Since the QCD EOS is not known (yet),
we shall demonstrate how to estimate $g_A$ using the parametric EOS
introduced in Ref.~\cite{Parotto:2018pwx}.
First, following Ref.~\cite{Pradeep:2022mkf}, we find
$Z$ by matching the leading singularity in the QCD EOS to that in the
Ising model:
\begin{multline}
  Z=\lim_{T,\mu\rightarrow T_c,\mu_c}\frac{c_pT}{n^2(T\xi)^{2-\eta}}
  =\frac{M_0T_c^4}{h_0n^2_c(T_c\xi_0)^{2-\eta}}\\
 \times  \left(\cot\alpha_1-\frac{s_c}{n_c}\right)^2
  \left[\frac{\sin\alpha_1}{w\sin(\alpha_1-\alpha_2)}\right]^2\,,
\end{multline}
where $w$, $\alpha_{1,2}$ and $\xi_0$ are parameters, defined in
Refs.~\cite{Parotto:2018pwx,Karthein:2021nxe}, which control the
orientation and strength of the critical point singularity located at
$T=T_c$ and $\mu=\mu_c$, with enthalpy given by
$w_c=n_c\mu_c+s_cT_c$. The same expression as in the square brackets
determines the width of the critical region \cite{Pradeep:2019ccv}.
The values of $M_0$ and $h_0$ are fixed in
Ref.~\cite{Parotto:2018pwx}.

Defining $\hat g_A$ so that $g_A\equiv\hat g_A\,
\sin\alpha_1/[w\sin(\alpha_1-\alpha_2)]$, we can use parameters in 
Refs.\cite{Parotto:2018pwx,Karthein:2021nxe} ($\mu_c=350\, \text{MeV}$,
$T_c=143.2\, \text{MeV}$, $\xi_0=1\, \text{fm}$) to estimate the values of the
couplings at zero momentum
($\bm p_A=\bm0$): $\hat g_{p,\bm0}\approx-3.1$,
$\hat g_{\pi,\bm0}\approx0.18$, $\hat g_{\bar
  p,\bm0}\approx5.5$.

The
approach in
Refs.~\cite{Stephanov:1998dy,Stephanov:1999zu,Stephanov:2008qz,Athanasiou:2010kw,Stephanov:2011pb,Bluhm:2016byc,Pradeep:2022mkf}
leading to Eq.~(\ref{eq:DG-h+fo}) leaves
not only the magnitude, but also
the sign of $g_A$  undetermined. While the overall sign can be
changed by redefining the critical field $\sigma$, the relative
sign of $g_A$ for different particles, or different momenta of the
same particle, i.e., different $A$, is not arbitrary and can be found
in the maximum entropy approach using Eq.~(\ref{eq:ga2}).  

Thus, we find that the critical mode
coupling to (low momentum) protons is opposite in sign from the coupling to
either pions or antiprotons. This can be traced back to the fact that
fluctuations of the number of protons contribute to the fluctuations
of the ratio $s/n$ with opposite sign from that of pions or antiprotons, since
pions contribute to the numerator, while protons (mostly) to the
denominator of the ratio.

Experimental implications of the changing sign of $g_A\sim P^m_A$ could be
studied by considering cross-species
correlators discussed in Ref.\cite{Athanasiou:2010kw} or 
correlations between particles with different momenta, i.e.,
$A\neq B$. In both cases
one would expect anticorrelation when the product $g_Ag_B\sim P^m_AP^m_B$
is negative.

\section{Conclusions}
\label{sec:conclusions}
Maximum entropy principle is widely used in many applications in
statistics, information theory, economics,
biology and bioinformatics, data science, computation, pattern
recognition, etc. Of course, thermodynamics itself is based on that
very principle. The thermodynamic state is, by definition, the state
of maximum entropy, i.e., the most likely ensemble of microscopic
states, given the known (i.e., measured) properties of the system,
such as total energy. The application to freezeout could be viewed as
answering the question of what is the most likely ensemble of
free-streaming particles after freezeout given the information about
the hydrodynamic conditions before the freezeout.

The key idea is that this information could include not only the values of
mean quantities but also of the hydrodynamic fluctuations (i.e.,
correlators $H_n$) out of equilibrium. These can be obtained, for example, from
a Hydro+ calculation
\cite{Stephanov:2017ghc,Rajagopal:2019xwg,Du:2020bxp,Pradeep:2022mkf},
or by solving full hydrodynamic fluctuation equations~\cite{An:2019osr,An:2019csj,An:2020vri}. Maximum entropy freezeout then determines the most likely ensemble
of  free-streaming final particles which matches all this available
information (equation of state and the predictions of hydrodynamics
with fluctuations).

Remarkably, the results are consistent with the picture, already
considered in the literature, of hadron gas coupled to fluctuating
fields inducing correlations. This not only corroborates
the picture, but provides a nontrivial insight into the entropic
origin of the correlations. Crucial for practical applications,
the maximum entropy approach provides information about the couplings
determining the magnitude of the correlations as well as the generalization
to non-Gaussian fluctuations in or out of equilibrium.

Obviously, it would be very interesting to implement this novel
approach in heavy-ion collision simulations
to explore the potential implications and to compare the
results with experimental data. In particular, the data from the Beam Energy
Scan at RHIC, whose results are being analyzed by the STAR
collaboration at this time. Such applications
are beyond the scope of this Letter and we defer these
investigations to future work.

\acknowledgments

We thank K.~Rajagopal and Y.~Yin  for helpful comments. This work is supported by
the U.S. Department of Energy, Office of Science, Office of Nuclear
Physics within the framework of the BEST Topical Collaboration and
grant No.\ DE-FG0201ER41195.

\newpage

\bibliographystyle{utphys}
\bibliography{refs_maxen}

\end{document}